AI.ƒ(T)

# The Insurability Frontier of AI Risk

*Mapping Threats to Affirmative Coverage, Silent Exposures, and Exclusions*

Alex Leung, Rex Zhang, Ervin Ling, Kentaroh Toyoda, SiewMei Loh

AIFT

**Abstract**

The rapid diffusion of agentic AI has created a new coverage problem for commercial insurance: some AI-mediated losses are now affirmatively insured, some create silent-AI exposure under legacy cyber, technology errors-and-omissions (E&O), directors-and-officers (D&O), employment practices liability (EPLI), crime, and media policies, and others are being actively excluded.

This paper maps that emerging boundary by coding 55 AI threat classes against 26 insurance products, endorsements, and exclusion regimes using public carrier materials and OWASP/MITRE threat catalogs. We identify a four-tier insurability frontier: affirmatively insured perils, silent-AI exposures, actively excluded perils, and perils outside conventional private insurance structures. Our coding measures publicly claimed positioning rather than executed contract wording; the headline statistics describe what carriers publicly state about coverage, not what would be paid in any specific claim.

Three patterns emerge. First, affirmative AI coverage is beginning to differentiate by primary risk emphasis: public materials often position Munich Re around model performance and drift, Armilla and parts of the Lloyd's market around hallucination and broader AI liability, Tokio Marine Kiln and CFC around IP and technology E&O concerns, Apollo ibott around emerging autonomous system liability, and Coalition around deepfake and AI-enabled cyber response. Second, legacy lines retain silent-AI exposure where AI is an instrumentality rather than the legal cause of loss. Third, foundation model concentration is the clearest genuinely novel insurability frontier because upstream model failure can correlate losses across many cedents at once; the relevant market design question is which insurability constraint each candidate structure relaxes, not merely which systemic risk template exists.

## I. INTRODUCTION

AI has become a coverage problem for commercial insurance. The same loss can now be described in several ways: as a cyber incident, a technology error, a professional services failure, a media or intellectual property claim, an employment or discrimination claim, a crime loss, or an AI-specific event. Existing policy forms were not drafted with that overlap in mind. At the same time, insurers are responding in opposite directions. A small affirmative market is offering AI-specific coverage, while parts of the broader liability market are adding AI exclusions or narrowing legacy wording.

That split creates a practical problem for buyers, brokers, reinsurers, and regulators. For a specific AI threat, for example hallucination, prompt injection, model drift, deepfake fraud, or an autonomous agent failure, it is often unclear whether the loss is expressly covered, creates silent-AI exposure under an older line, is excluded by endorsement, or falls outside conventional private insurance structures. The ambiguity matters because AI risks are not one category. Some are ordinary liability risks with a new instrumentality; some are cyber or technology risks reframed through AI systems; some raise intentional act, moral hazard, or aggregation problems that standard insurance tools handle poorly.

This paper addresses that problem by translating AI security threats into insurance coverage terms. We build a 55-threat by 26-product matrix that maps published AI threat classes against affirmative AI products, legacy insurance lines, and AI-specific exclusion regimes. The matrix is based on public carrier and partner materials, regulatory and form-filing materials where available, trade press reporting, and the OWASP and MITRE threat catalogs current to May 2026. The goal is not to predict the outcome of any particular claim. Actual coverage depends on executed wording, jurisdiction, facts, retentions, sublimits, exclusions, and other insurance provisions. Instead, the goal is to make the emerging market structure visible at the threat level.

Because the evidence base is public, the matrix measures observable market positioning rather than claim payment outcomes. An affirmative code means that public materials expressly identify the peril or use wording broad enough to capture it under the paper's decision rules; it does not mean that every claim involving that peril would be paid.

The paper makes three contributions. First, it provides a replicable codebook for classifying AI-related insurance positions as affirmative, silent, excluded, or not publicly addressed. Second, it uses that codebook to identify a four-tier insurability frontier: perils that are affirmatively insured; perils that create silent-AI exposure under legacy lines; perils that are actively excluded; and three structurally distinct boundary cases that we treat as Tier 4 subtypes: architectural exploitability (the lethal trifecta), the longstanding doctrinal exclusion of intentional acts (AI-washing), and systemic loss correlation across cedents (foundation model concentration). We argue that only the third is a genuinely novel insurability frontier; the first calls for controls before coverage, and the second is a familiar problem in new clothing. Third, it isolates foundation model concentration as the clearest systemic aggregation problem in the corpus, distinct from more familiar AI perils such as hallucination or model drift.

The central finding is that the AI insurance market is not merely expanding or contracting. It is sorting AI risk by peril. Public materials suggest that some carriers are visibly positioned around model performance and drift, some around hallucination and AI liability, others around IP defense, autonomous system liability, or deepfake response. At the same time, legacy lines retain silent-AI exposure, and exclusion endorsements are reducing that ambiguity by removing coverage rather than clarifying it. The resulting frontier is useful for placement strategy, reinsurance aggregation analysis, regulatory disclosure, and a standing liaison between AI security taxonomy projects and the affirmative AI cover market.

### *A. The coverage problem*

The coverage problem begins with causation. An AI-mediated loss can arise from a model output, an adversarial prompt, a compromised tool, a human decision to rely on an AI system, a vendor failure, or a downstream injury caused by an automated action. Each framing points to a different insurance line. A hallucinated legal citation may look like professional negligence; a poisoned retrieval source may look like cyber intrusion or technology E&O; a synthetic executive video may look like crime, social engineering, or cyber response; and a foundation model failure may produce correlated claims across many insureds at once.

That framing problem is what we call a silent-AI exposure: an AI-mediated loss falling under a non-AI policy that did not clearly contemplate AI as a cause, instrumentality, or excluded peril. Silent-AI exposure is analogous to silent cyber, but it is not identical. Silent cyber usually asks whether a cyber event is embedded inside a non-cyber loss. A silent-AI exposure often asks a harder question: whether AI is the cause of loss, the tool used to create the loss, the product that failed, or merely part of the factual background.

### *B. What this paper does*

We organize the problem in two dimensions. The first dimension is the threat: 55 AI-related perils drawn from LLM application risk, agentic AI risk, agent infrastructure, classical machine learning risk, AI operations and supply chain, human misuse, systemic aggregation, and agentic-skill risk. The second dimension is the insurance response: 26 affirmative products, legacy lines, and exclusion regimes. Each cell receives one of four codes: affirmative, silent or gray, excluded, or no public position.

This design lets us ask a concrete question: for each AI threat, where does the public insurance market currently appear to place the risk? The answer is a map rather than a legal opinion. The map shows which threats are moving into affirmative AI cover, which remain dependent on contested legacy wording, which are being excluded, and which are better addressed through architecture, governance, vendor contracts, or market level aggregation tools.

### *C. Roadmap*

The rest of the paper proceeds in four steps. Sections II to V define the insurability framework, source base, threat catalog, and insurance product catalog; Sections VI and VII present the matrix results and systemic aggregation analysis; Sections VIII to X discuss implications, limitations, and conclusions.

## II. RELATED WORK AND INSURABILITY FRAMEWORK

### *A. Insurability criteria*

The notion of an insurability frontier predates AI by decades. The classical criteria associated with Berliner (1982), namely randomness of loss occurrence, maximum possible loss, average loss per event, loss exposure, information asymmetry, insurance premium, cover limits, and public policy compatibility, have since been applied to cyber, terrorism, pandemic, and climate risk. We compress Berliner's eight criteria into six by treating average loss and loss exposure as part of assessability, treating cover limits as part of bounded maximum loss, and treating public policy compatibility chiefly through the intentional act / moral hazard criterion. Subsequent scholarship has refined these for novel risks in ways directly relevant to AI: Romanosky and colleagues have used cyber insurance pricing and claims data to study cyber insurability empirically [49]; Talesh has examined how cyber insurers function as de facto regulators by translating contested risks into underwriting standards [50]; and Baker has theorized the broader role of insurance in governing emerging technological risk [51]. We draw on this lineage in characterizing AI as a novel risk problem rather than a pure Berliner exercise. We use a compact six-criterion form throughout this paper:

1. randomness/fortuity of loss;
2. assessability of probability and severity;
3. loss independence (so the law of large numbers operates);
4. maximum possible loss bounded;
5. economic feasibility (premium acceptable to cedent and insurer); and
6. absence of intentional act / moral hazard.

We apply these criteria in Section VI and again to the systemic aggregation peril in Section VII.

### *B. Silent cyber as the analog*

The concept of silent cyber, namely cyber-mediated losses falling under non-cyber policies that did not contemplate cyber as a peril at inception, emerged after NotPetya (2017) and was central to Lloyd's market bulletins requiring affirmative or excluded cyber wording on all property treaty business by 2020. We use silent-AI exposure to denote AI-mediated losses falling under non-AI policies (CGL, D&O, E&O, EPLI, crime, media liability) that contemplated neither AI as cause nor AI as instrumentality at inception. The rapid 2025 to 2026 development of both affirmative AI products and AI-specific exclusions tracks the silent-cyber bifurcation closely, and we draw on that precedent throughout.

### C. Recent insurance industry literature

The recent insurance industry literature on AI risk has three strands. First, performance guarantee work centered on Munich Re's aiSure product, which has insured against model drift as the headline peril since 2018 [5, 6, 39]. Second, affirmative liability work catalyzed by Armilla's Lloyd's-backed launch in April 2025 and AIUC's emergence from stealth in July 2025 [16, 26, 28]. Third, the exclusion-driven response from the broader market documented by the Financial Times and trade publications [9, 33, 37, 38, 47]. Aon's Kevin Kalinich, in remarks reported directly by the Financial Times, articulated the systemic argument that motivates Section VII: carriers may absorb a very large single-insured AI loss, but cannot readily price an upstream AI-provider mistake that produces "a systemic, correlated, aggregated risk" across many insureds [9]. The Geneva Association's 2025 work on AI in insurance and the Munich Re whitepaper series describe the macro view but do not provide threat level mappings [5, 23].

### D. AI security literature

The AI security literature has converged on a small set of authoritative threat catalogs: the OWASP Top 10 for LLM Applications for application-layer risk [1]; OWASP Top 10 for Agentic Applications for agentic risk [3]; the OWASP Agentic Skills Top 10, an incubator project current as of April 2026, for the skill-layer risk that emerged with agent-skill ecosystems [11, 40, 41]; MITRE ATLAS v5.1.0 (November 2025) with 16 tactics and 84 techniques [4]; and Simon Willison's lethal-trifecta formulation, which identifies the architectural pattern (private data, untrusted content, and external communication) as a deterministic root of prompt injection exploitability [7]. Beurer-Kellner et al. (2025) provide design patterns for defending against prompt injection [12], and Meta's LlamaFirewall and OpenGuardrails projects represent open source guardrail efforts that the authors describe as experimental [44, 45]. To the best of our knowledge, no published work systematically crosses these threat catalogs against the affirmative/silent/excluded coverage status of named insurance products at the granularity of this paper.

## III. METHODOLOGY

### A. Sources and evidentiary tiers

We assembled a corpus from three categories of sources, each with different evidentiary weight.

Tier-A (carrier-published). Carrier or partner-published product pages, press releases, and public facing product descriptions for Armilla, AIUC, Testudo, Coalition, Axa XL, Munich Re, Munich Re HSB, Vouch, Apollo ibott, and the Marsh+ibott Autonomous Vehicle Insurance Program (AVIP) for Uber [16, 21, 24-32, 39]. Where a Tier-A source describes coverage in marketing grade language (e.g., "covers AI hallucination"), we treat the cell as affirmative.

Tier-B (regulatory and standards filings). Verisk's CG 40 47/48 generative AI exclusion endorsements [21, 22]; W. R. Berkley's PC 51380 where filing detail is available [37]; and regulator published AI standards and circulars (including the NAIC Model Bulletin on AI, NY DFS Circular Letter No. 7, the EU AI Act, and Colorado AI Act SB 24-205) [23, 46]. Tier-B sources are public record filings or regulator materials; for these we coded the cell directly from the filing text where available.

Tier-C (trade press and analyst commentary). For carrier-specific exclusion filings whose underlying state-by-state forms are not always public, we triangulated across the Financial Times, Business Insurance, Reinsurance News, American Banker, S&P Global Market Intelligence, the Insurance Intel Substack, and Slipcase [9, 33, 35-38, 47]. Evercore ISI's market exposure ranking (Chubb, AIG, Zurich, AXA XL, Allianz as "most exposed") was reported by Reinsurance News in December 2025 [47]. Tier-C cells are coded conservatively: where a single trade press source asserts a coverage position not corroborated by Tier-A or Tier-B, we coded the cell as silent rather than affirmative or excluded.

### B. Threat catalog

We cataloged 55 threats grouped into seven clusters: LLM/GenAI application threats (T-01 to T-10, plus T-43 AI-washing); agentic AI threats including the OWASP Agentic Top 10 ASI01 to ASI10 mapped to T-11 to T-20, plus T-40 (lethal trifecta), T-41 (browser-agent hijack), T-42 (tool description injection), and T-44 (AI worms); the MCP/agent infrastructure cluster (T-21 to T-24); classical ML/deep-learning threats (T-25 to T-30); AI ops and supply chain (T-31 to T-34); people and misuse (T-35 to T-39); systemic and aggregation (T-45); and the OWASP Agentic Skills Top 10 incubator project mapped to T-46 to T-55. Each threat carries source citations, framework references (OWASP IDs, MITRE ATLAS techniques, NIST AI RMF), and where available real-world cases. Several threats overlap substantially when viewed through a coverage lens. For example, T-01 (prompt injection) and T-02 (indirect prompt injection) almost always co-occur for coverage purposes; T-44 (AI worms) is largely a propagation vector for T-02; and the MCP cluster (T-21 to T-24) overlaps with several AST10 entries (T-47, T-48, T-51) on supply-chain and skill-isolation themes. We retain the fine-grained taxonomy because it preserves traceability to the underlying OWASP and MITRE catalogs that carriers and underwriters actually reference, and because two threats that look similar in security terms can map to different policy lines (e.g., social engineering vs. cyber). The cost is that the "55 threat" headline reflects taxonomic granularity as well as coverage distinct perils.

### C. Insurance product catalog

We cataloged 26 named insurance items in three categories: 14 affirmative AI products and endorsements (I-01 through I-14); six pre-existing lines that may create silent-AI exposure (I-15 standard cyber, I-16 tech E&O, I-17 D&O, I-18 EPLI, I-19 crime/social engineering, I-20 media liability); and six exclusion or expected exclusion regimes (I-21 through I-26, ranging from the Verisk CGL endorsement to absolute multi-line carrier exclusions to design-professional E&O carve-outs). Inclusion criteria for current affirmative or exclusion coding are Tier-A or Tier-B public evidence. I-26 is retained as a Tier-C expected-exclusion signal, but its affected cells are coded silent rather than excluded unless filed wording becomes public. As a result, the product catalog captures market signals while the matrix distinguishes current public filings from forecast renewal-cycle positions.

### *D. Coding scheme and codebook*

We coded each (threat, product) cell with one of four values:

**A. Affirmative.** The product, as described in Tier-A or Tier-B sources, expressly covers losses arising from this threat. The peril is named (e.g., "hallucination") or the cause-of-loss language captures it directly (e.g., "AI Model Error Liability").

**S. Silent / gray.** The product does not affirmatively cover the threat but does not exclude it either. Coverage in any actual claim depends on contested wording, the legal characterization of the loss, sublimits and retentions, or the interaction of the AI peril with a non-AI peril (e.g., whether prompt injection of a deployed LLM constitutes "unauthorized access to a Computer System" for cyber-policy purposes).

**X. Excluded.** The product expressly excludes the threat, typically via a named-peril exclusion (e.g., Verisk CG 40 47's "arising out of generative artificial intelligence") or by virtue of the product itself being an exclusion endorsement applied to an underlying line.

· **Not applicable / no public position.** The product does not address this threat in any direction; either the product line is structurally unrelated (e.g., Coalition's deepfake endorsement vs. model drift) or insufficient public evidence exists to assign a position.

### *E. Decision rules for ambiguous cells*

We adopted four decision rules to reduce coder discretion in ambiguous cells. (1) Marketing grade naming rule: if the carrier's public product description names a peril ("hallucination", "bias", "IP infringement"), we coded the cell affirmative even where the executed policy wording would be required to confirm a claim. (2) Cause-of-loss capture rule: where the carrier's wording uses a generic AI-error class ("AI Model Error", "AI Agent Failure") that plausibly captures the threat, we coded affirmative; otherwise silent. (3) Exclusion-precedence rule: where a product is itself an exclusion endorsement applied to an underlying line (I-21 through I-26), we coded the cell as excluded for any threat the endorsement's "arising out of AI" wording would capture, regardless of whether the underlying line would otherwise have responded. (4) Single-source rule: where the only evidence is a single Tier-C trade press citation, we coded the cell silent, not affirmative.

Rule (1) is in unavoidable tension with the silent-cyber thesis that motivates this paper: the entire point of distinguishing affirmative from silent coverage is that marketing language and contract wording can diverge. Coding a cell as affirmative on the basis of a named peril in a product page therefore measures publicly claimed coverage, not coverage as it would be applied in a particular claim. We adopt the rule for replicability, since public materials are observable and reproducible while executed wording is generally not, but we report headline statistics throughout the paper as descriptions of public market positioning rather than predictions of claim outcomes. A more conservative coding that required wording grade evidence would yield substantially fewer affirmative cells; we treat the marketing grade reading as an upper bound on observable affirmative cover and flag this explicitly when reporting aggregate counts.

### *F. Coding process and inter-rater reliability*

Each cell was coded independently by three human coders working from the same evidence pack and decision rules described in Sections III.D and III.E. Coders were assisted by large language models for source retrieval, candidate rule application, and consistency checks across analogous cells, but the final code in each cell reflects a human decision; LLM outputs were treated as a research aid rather than a coding authority. Models used for that assistance layer were Anthropic Claude Opus 4.7 and OpenAI GPT-5.5. To preserve coder independence, no LLM-generated recommendation was treated as a coder vote, and human coders finalized each cell before reviewing analogous cell consistency reports. Where the three coders disagreed, the cell was flagged for adjudication and resolved by group discussion against the evidence pack; remaining contested cells were coded toward the more conservative end of the dispute (e.g., affirmative downgraded to silent, or exclusion downgraded to silent or no-position depending on whether any non-exclusion public evidence remained).

### *G. Limits of the coding*

Three limits warrant immediate disclosure. First, for several carrier-specific exclusion filings (notably the multi-line filings under I-24), the underlying state-by-state forms are not always public; we rely on trade press summaries. Second, the OWASP Agentic Skills Top 10 remains an incubator project as of May 2026; the AST cluster (T-46 to T-55) is therefore the most volatile portion of the threat catalog. Third, even with three-coder adjudication, the matrix codes a single best evidence interpretation per cell rather than a distribution across plausible coverage arguments; future work could extend this to a probabilistic coding that reflects coder confidence and source tier weighting. The full per-cell evidence index is reproduced in supplementary material as a numbered source map.

## IV. THE THREAT CATALOG

Table I summarizes the eight threat clusters. We highlight five threats that operate as the structural anchors of our subsequent analysis.

T-02 Indirect prompt injection. Hidden adversarial instructions in documents, web pages, emails, or RAG sources. EchoLeak (CVE-2025-32711) was reported as a zero-click prompt-injection vulnerability in Microsoft 365 Copilot [42]. A cluster of 2026 disclosures involving AI-enabled work tools and OpenAI's December 2025 Atlas hardening work [2] show that indirect prompt injection remains a dominant attack surface for LLM-mediated workflows.

T-29 Model drift. Real-world data distribution diverges from training; performance silently decays. This is the headline use case for Munich Re's aiSure product, which has operated since 2018 [6, 39].

T-35 Deepfake fraud. Synthetic audio, video, or image content impersonates executives, customers, or counterparties. The Arup $25M Hong Kong wire transfer scam (2024) cloned a CFO's image and voice in a video conference. Coalition's Deepfake Response Endorsement, announced globally on 9 December 2025, is a dedicated first-party response product [30].

T-40 Lethal trifecta. Simon Willison's architectural risk formulation: any agent with private data access plus untrusted content exposure plus external communication is structurally vulnerable to data theft via prompt injection [7, 19]. Filters reach approximately 95% accuracy on known patterns; in security critical use, "95% is a failing grade." The only deterministic defense is to architecturally cut one leg.

T-45 Foundation model concentration. Enterprise AI is concentrated among a small number of foundation model providers. As Aon's Kalinich notes, traditional liability underwriting assumes loss independence, an assumption that may no longer hold when one upstream provider's failure can simultaneously affect thousands of cedents [9]. We develop this as the paper's focal systemic aggregation case in Section VII.

TABLE I: Threat clusters and IDs used in this paper.

| Cluster | Count | IDs (representative) | References |
|---|---|---|---|
| **LLM / GenAI application** | 11 | T-01 prompt injection, T-04 hallucination, T-10 IP infringement, T-43 AI-washing | OWASP LLM Top 10 |
| **Agentic AI** | 14 | T-11 goal hijack, T-20 rogue agents, T-40 lethal trifecta, T-41 browser hijack, T-44 AI worms | OWASP ASI01-10 |
| **MCP / agent infra** | 4 | T-21 MCP compromise, T-22 unauth servers, T-24 OAuth cascade | OX Security, Vulcan MCP-38 |
| **Classical ML / DL** | 6 | T-25 evasion, T-26 poisoning, T-29 drift, T-30 bias | MITRE ATLAS, NIST AI RMF |
| **AI ops & supply chain** | 4 | T-31 compromised foundation model, T-33 lack of provenance | OWASP LLM03, CISA SBOM-AI |
| **People / misuse** | 5 | T-35 deepfake fraud, T-37 shadow AI, T-39 regulatory non-compliance | FBI IC3, EU AI Act |
| **Systemic / aggregation** | 1 | T-45 foundation model concentration | Aon, Geneva Association |
| **Agentic skills** | 10 | T-46 malicious skills, T-47 supply chain, T-51 weak isolation | OWASP AST01-10 (in incubation) |

## V. THE INSURANCE-PRODUCT CATALOG

Table II summarizes the 26 products. Six observations frame the rest of the paper.

First, the affirmative AI cover market is dominated by managing general agents (MGAs), namely Armilla, AIUC, Testudo, and Vouch, writing on Lloyd's-backed or syndicate paper, plus the established Munich Re aiSure program. Second, the affirmative products differ markedly in structure: Munich Re aiSure is publicly described as a contractual liability arrangement backing a Performance Guarantee that the Insured offers to its customers [6, 39]; Armilla and AIUC are liability policies with affirmative AI triggers; Apollo ibott Syndicate 1971 wraps autonomous-system exposure [31].

Third, the exclusion-side filings span a wide range, from Verisk's targeted generative AI wording to W. R. Berkley's broader multi-line exclusion. Fourth, the silent-AI zone in existing lines remains large: standard cyber, tech E&O, D&O, EPLI, crime, and media liability all contain wording that may or may not respond to AI-mediated losses depending on the incident. Fifth, several carriers occupy more than one row in the taxonomy. Sixth, the threat-to-market map that emerges (Table III) suggests specialization by peril rather than broad uniform coverage: Munich Re is most visibly positioned around model drift, Armilla/Chaucer/Hiscox around hallucination, Tokio Marine Kiln/CFC around IP defense, Apollo ibott around autonomous and agentic liability, and Coalition around deepfake response.

TABLE II: Insurance products, endorsements, and exclusions in the corpus. “Aff.” = affirmative AI cover; “Excl.” = exclusion endorsement; “Existing” = pre-existing line with possible silent-AI exposure or carve-outs.

| ID | Visible market positioning | Product / Endorsement | Carrier(s) | Status / notes |
|---|---|---|---|---|
| I-01 | Aff. liability | Armilla AI Liability | Armilla MGA; Chaucer; Axis Capital | Apr 2025 launch; raised to $25M Jan 2026 |
| I-02 | Aff. performance | Munich Re aiSure (performance guarantee) | Munich Re; Mosaic syndication | Operating since 2018; mid-market sizing per public materials |
| I-03 | Aff. agent liability | AIUC AI Liability ($50M) | AIUC (Artificial Intelligence Underwriting Company) | Out of stealth Jul 2025; tied to AIUC-1 audit |
| I-04 | Aff. mid-market | Testudo AI Liability | Testudo MGA, Lloyd’s-backed; Lloyd’s Lab Cohort 14 | Underwriting commenced Jan 2026 |
| I-05* | Aff. cloud-bundled cyber | Google Cloud Risk Protection Program (AI endorsement) | Beazley; Chubb; Munich Re (founding 2021); Munich Re Specialty / HSB | RPP launched 2021; AI insurance endorsement added May 2025; layered atop Google Cloud 2023 IP indemnification [54, 56] |
| I-06 | Aff. in cyber | Coalition Active Cyber, AI endorsement | Coalition | In base policy from 2025; SEC cyber disclosure cover bundled |
| I-07 | Aff. response | Coalition Deepfake Response Endorsement | Coalition | Announced globally 9 Dec 2025 |
| I-08 | Aff. in cyber | Axa XL Generative AI endorsement | Axa XL | Introduced October 2024 |
| I-09 | Aff. SMB | HSB AI Liability for SMB | Munich Re HSB | Launched mid-March 2026 |
| I-10 | Aff. tech-startup | Vouch AI Insurance + Hiscox/Corix | Vouch (specialty MGA); Corix (Hiscox division) | Original launch 2024; “Vouch Horizon” AI risk team |
| I-11 | Aff. warranty | Armilla AI Product Warranty | Armilla | Vendor differentiator in enterprise sales |
| I-12 | Aff. hallucination | Hiscox / Chaucer Hallucination Liability | Hiscox; Chaucer (Lloyd’s syndicates) | Active 2025-2026 market positioning |
| I-13 | Aff. IP defense | Tokio Marine Kiln / CFC IP Defense (AI extension) | Tokio Marine Kiln (TMK); CFC Underwriting | Lloyd’s specialty; established markets extending to AI IP |
| I-14 | Aff. autonomous | Apollo ibott Autonomous / Agentic Liability | Apollo Group (Skyward), Syndicate 1971 “ibott” | Marsh+ibott AVIP for Uber announced 12 Mar 2026 |
| I-15 | Existing | Standard cyber liability (1st & 3rd party) | AIG; Chubb; Beazley; Travelers; AXA XL; Allianz; Zurich; etc. | Silent-AI exposure problem; carriers narrowing wording in 2026 |
| I-16 | Existing | Technology Errors & Omissions | Tokio Marine HCC; AIG; Beazley; Hiscox; etc. | Often excludes or under-insures in-house AI; gap targeted by Armilla and Vouch |
| I-17 | Existing | Directors & Officers liability | All major D&O carriers | Aggressive AI exclusions emerging; SEC AI-washing settlements active |
| I-18 | Existing | Employment Practices Liability | All major EPLI carriers; Chaucer / Beazley lead AI-driven third-party liability | Underwriters demanding documented bias controls; coverage tightening |
| I-19 | Existing | Crime / Social Engineering | All major crime carriers | Active gray zone on whether AI-generated content counts as “direct” communication |
| I-20 | Existing | Media / Multimedia liability | Beazley; Chubb; Hiscox; Travelers; Axis | AI sublimits and carve-outs emerging; relevant for ad-tech and publishers |
| I-21 | Excl. | Verisk ISO CG 40 47 / CG 40 48 / CG 35 08 | Adoptable by any US carrier using ISO forms | Effective 1 Jan 2026; matrix codes only threats directly captured by public generative-AI wording |
| I-22 | Excl. | W. R. Berkley PC 51380 (absolute AI exclusion) | W. R. Berkley | Filed late 2025; broader than Verisk; D&O / E&O / fiduciary |
| I-23 | Excl. | Berkley absolute AI exclusion (CGL / multi-line) | Berkley (separate filing) | Independent filing pre-Verisk; bars AI use, deployment, or development |
| I-24 | Excl. | Carrier-specific AI exclusion filings | AIG; Great American; Cincinnati Financial; Frederick Mutual; Philadelphia | Filed 2025-2026; coded narrowly where public filing summaries support direct threat capture |
| I-25 | Excl. | Design Professional E&O AI exclusions | Design-pro specialty carriers | Effective Jan 2026 cycle; cites Stanford 20-33% / 58-88% hallucination rates |
| I-26 | Expected excl. / Tier-C | Industry-wide GenAI liability exclusions (expected) | Chubb; AIG; Zurich; AXA XL; Allianz (Evercore ISI most exposed list) | Expected renewal cycle issue; Tier-C-only signals are coded S, not X, absent filed wording |

* Note. I-05 is positioned in the “affirmative” category at the product level under Tier-A material describing workload-class coverage, but its threat-by-threat cells are coded “silent” under the marketing grade naming rule (§3.5) because no individual perils are named in carrier-published materials.

TABLE III: Threat-to-market positioning crosswalk. Markets are listed in approximate order of public market positioning per carrier filings, market intelligence, and trade-press reporting [9, 16, 28, 31, 47, 48].

| Peril | Visible market positioning | Cover type |
|---|---|---|
| **Hallucination** | Armilla / Chaucer / Hiscox | Affirmative liability / warranty |
| **Model drift** | Munich Re (aiSure) / Armilla | Performance guarantee structure |
| **Algorithmic bias** | Chaucer / Beazley | Third-party liability |
| **IP infringement** | Tokio Marine Kiln / CFC | IP defense coverage |
| **Agentic failure** | Apollo ibott Syndicate 1971 | Autonomous / agentic liability |
| **Deepfake response** | Coalition | First-party response endorsement |

## VI. THE THREAT × INSURANCE MATRIX

### *A. Aggregate statistics*

The headline matrix is 55 threats by 26 products = 1,430 cells. In the current published matrix, 243 cells (17.0%) are coded affirmative, 127 (8.9%) are silent / gray, 91 (6.4%) are excluded, and 969 (67.8%) are inapplicable or have insufficient public evidence to assign a position. Concentrating on the 461 cells where any position is taken, a more meaningful denominator, 52.7% are affirmative, 27.5% are silent, and 19.7% are excluded. Affirmative public market positions therefore modestly outnumber silent and excluded positions combined, consistent with active bifurcation rather than a settled equilibrium. We re-emphasize that these counts measure publicly claimed positioning under the marketing-grade naming rule (Section III.E), not coverage as it would be applied in any specific claim. The "52.7% affirmative" figure should therefore be read as: of the cells where the public market takes any visible position, slightly more than half are positions of claimed cover. A wording grade recoding requiring executed policy evidence would yield a substantially smaller affirmative share. The bifurcation finding is robust to this caveat at the directional level, but the specific point estimates are upper bounds on observable affirmative cover.

Per-product coverage breadth varies widely under the coding rules. The figures in this paragraph are direct column counts from Appendix A. Among affirmative products, AIUC is coded with 44 A cells, Coalition's AI cyber endorsement with 41, Armilla with 30, and Testudo with 22, reflecting broad public AI-agent-failure, AI-model-error, and cyber-plus-AI positioning. Apollo ibott is coded with 16 A cells focused on autonomous and agentic action. Munich Re aiSure, by contrast, is coded with 12 A cells, a narrower but structurally precise scope, because the product is publicly positioned around a contractual performance guarantee rather than open-ended liability [6, 39]. On the exclusion side, direct Appendix A counts show W. R. Berkley PC 51380 (I-22) with 39 X cells and design-professional E&O exclusions (I-25) with 33 X cells. The Verisk generative AI exclusion column (I-21) and carrier-specific multi-line filings (I-24) are coded narrowly at 3 X cells each under the current evidence rule; a broader CG 40 47/48 reading would require a corresponding matrix expansion rather than a prose-only claim. I-26 is retained as an expected-exclusion signal, but its nine Tier-C-only cells are coded S rather than X. Per-threat coverage breadth also varies sharply: IP infringement (T-10) and evasion / KYC (T-25) are the most affirmatively coded threats at 9 A cells each, followed by hallucination (T-04), sensitive information disclosure (T-03), and data poisoning (T-26) at 8 A cells each. Shadow AI (T-37) draws no affirmative covers and multiple exclusions; AI-washing (T-43) is predominantly excluded or unaddressed, with no affirmative cells after the Appendix A re-audit.

### *B. The four-tier insurability frontier*

Aggregating the matrix produces a four-tier insurability frontier, with Tier 4 itself comprising three structurally distinct boundary cases (4a, 4b, 4c) that we develop separately. Table IV summarizes the structure; Sections VI.C to VI.F develop each tier. I-26 is grouped with the exclusion family by market intent, but its current cells remain in the Tier 2 silent-AI exposure zone because the public evidence is forecast-oriented rather than filed wording.

TABLE IV: The four-tier insurability frontier for AI risk.

| Tier | Insurability criteria* status |
|---|---|
| 1: Affirmatively insured | All six criteria substantially met |
| 2: Silent-AI exposure | Criterion (ii) assessability and (vi) intentional act ambiguous |
| 3: Actively excluded | n/a; exclusion is a market response, not an insurability finding |
| 4: Outside conventional/private insurance | Criteria (i), (iii), or (vi) structurally stressed |

* Note. Criteria: (i) randomness/fortuity of loss; (ii) assessability of probability and severity; (iii) loss independence; (iv) maximum possible loss bounded; (v) economic feasibility; and (vi) absence of intentional act / moral hazard.

### *C. Tier 1: Affirmatively insured perils*

Hallucination (T-04), copyright/IP infringement (T-10), sensitive information disclosure (T-03), model drift (T-29), data poisoning (T-26), evasion (T-25), and bias (T-30) are coded as affirmatively covered by multiple public-market products. Hallucination has the broadest market response in the corpus: Armilla, AIUC, Munich Re aiSure, Testudo, Vouch+Hiscox/Corix, Hiscox/Chaucer, Coalition's AI cyber endorsement, and Armilla's AI Product Warranty are all coded as affirmative at the threat level. Against the six insurability criteria, Tier 1 perils score reasonably well: loss events can be treated as individually random conditional on deployment volume; severity is at least partially assessable; events are often independent at the per-deployment level; maximum loss is bounded by contract or claim limit; pricing is feasible because multiple markets quote the peril; and intentional act exposure is usually external to the insured rather than inherent in the peril. A Mata v. Avianca-style hallucinated citation incident illustrates the placement problem: hallucination can be affirmatively coded at the AI-liability layer while the professional liability or tech E&O response remains wording dependent unless AI use is expressly addressed.

### *D. Tier 2: Silent-AI exposure perils*

Cost-DoS (T-08), shadow AI (T-37), tool description injection (T-42), and lack of provenance (T-33) sit predominantly in the silent-AI exposure zone. Standard cyber and tech E&O may respond depending on framing; affirmative AI products mostly do not address these specifically. The lethal trifecta itself (T-40) carries no A or X codes anywhere in Appendix A; affirmative AI products and standard cyber/E&O lines are coded silent (S), while many other products are unaddressed (·). This is a notable observation, since the trifecta is what creates the exploitability of T-01, T-02, and T-11. Tier 2 perils typically fail criterion (ii) assessability of probability and (vi) absence of intentional act in mixed ways: shadow AI involves voluntary employee misuse; cost-DoS attacks are intentional but on the third-party side; tool description injection is a hybrid of supply chain and architectural cause. The operational pattern is visible in two recurring examples. A synthetic executive video may be affirmatively covered for response costs under a dedicated deepfake endorsement, while the funds transfer loss still turns on crime/social-engineering wording. A Drift-style OAuth cascade can be a cyber event, an AI-agent-failure event, and an excluded CGL event at the same time; the placement issue is allocation across primary/excess, anti-stacking, and AI-exclusion wording rather than whether the event is simply cyber or AI.

### *E. Tier 3: Actively excluded perils*

Under the conservative public evidence rule used here, Verisk CG 40 47/48 (I-21) is coded as excluding only the three threats directly captured by the published generative AI wording, rather than the broader catalog the endorsement could capture under a literal reading. Under a broad "arising out of" reading, CG 40 47/48 could capture a substantially larger fraction of the catalog; the matrix codes only directly captured threats to maintain a public evidence floor, but practitioners should expect the operational reach of the endorsement to be wider than the cell count suggests. The largest active-exclusion columns are W. R. Berkley's PC 51380 (I-22), coded with 39 X cells, and design-professional E&O exclusions (I-25), coded with 33 X cells. Carrier-specific multi-line filings (I-24) are also coded narrowly, with 3 X cells, because the publicly available filing summaries do not support a threat-by-threat expansion to the full catalog. I-26 is treated differently: expected GenAI exclusions among large incumbent carriers remain a developing renewal-cycle issue rather than a settled universal market position [9, 47], so its nine affected cells are coded S under the single-source rule rather than X. The design-professional E&O exclusions target AI-generated specifications, code, and drawings, citing research on legal and general-purpose AI hallucination rates [34]. Tier 3 is therefore a market structure choice rather than a pure Berliner failure: many excluded perils are technically insurable elsewhere in the matrix, but specific carriers or lines are choosing not to absorb them on legacy forms.

### *F. Tier 4: Three structurally distinct boundary cases*

Three threats sit outside the affirmative / silent / excluded structure of Tiers 1 to 3, but for structurally different reasons that we treat as separate subtypes. Lumping them together as a single "outside conventional insurance" tier obscures more than it reveals: the appropriate response, the binding insurability criterion, and the policy implication differ in each case.

Tier 4a: Architectural exploitability (controls, not coverage)

T-40 (lethal trifecta) stresses criterion (i) randomness/fortuity because the root cause is architectural and conditional on configuration, not purely stochastic. The first insurance economic response is therefore configuration change, namely cutting one of the three legs of the trifecta, and the second is underwriting discipline around architecture, permissions, and monitoring. We list this as a Tier 4 case because current public products do not appear to price configuration risk directly; that does not mean insurance can never respond to downstream losses. The implication for risk managers is that procurement of affirmative AI cover does not substitute for architectural mitigation; the implication for underwriters is that trifecta posture should be a risk selection and conditions issue before it is treated as a standalone priced peril.

Tier 4b: Intentional act exclusions (a familiar problem)

T-43 (AI-washing) stresses criterion (vi) absence of intentional act. AI-washing is, definitionally, the issuer's voluntary misrepresentation; standard intentional acts and known-falsity exclusions in D&O and E&O policies are therefore central to the analysis, and SEC enforcement actions against Delphia and Global Predictions established the doctrinal contours [25]. We caution, however, that this is an old frontier rather than a new one: intentional misrepresentation has always sat outside private liability cover, and the role of AI here is largely as the subject matter of the false statement, not as a new feature of insurability theory. Appendix A now codes no affirmative cells for T-43 because no public Apollo ibott material affirmatively claims coverage for securities law AI-washing; the prior agentic misstatement reading was therefore too broad for the paper's public-evidence rule. The Tier 4 listing is mainly to flag that buyers should not expect D&O to backstop AI-washing exposure that turns on knowing or reckless conduct, and that AI-specific carve-outs in D&O are largely codifying existing doctrine rather than introducing new restrictions.

Tier 4c: Systemic loss correlation (novel insurability frontier)

T-45 (foundation model concentration) stresses criterion (iii) loss independence at the portfolio level: a foundation model provider's failure can produce simultaneous claims across the cedent base, defeating the law of large numbers. Appendix A codes AIUC, Apollo ibott, and standard cyber as silent for T-45 at the individual policy level; these per-cell silent positions mean only that a single insured's downstream claim could create an allocation question. No public product in the corpus covers aggregation across cedents. Of the three Tier 4 subtypes, only this one is genuinely novel as an insurability problem: it is not addressable by controls alone (Tier 4a), nor a long-standing exclusion category (Tier 4b), but a structural correlation problem that requires a market level rather than policy level response. We develop this case in Section VII.

TABLE V: Application of insurability criteria to representative perils. ✓ = criterion substantially met; ~ = criterion ambiguous or partially met; ✗ = criterion structurally violated.

| Peril | (i) Rand. | (ii) Assess. | (iii) Indep. | (iv) Bound. | (v) Econ. | (vi) Intent | Net status |
|---|---|---|---|---|---|---|---|
| **T-04 Hallucination (Tier 1)** | ✓ | ✓ | ✓ | ✓ | ✓ | ✓ | Insurable |
| **T-29 Model drift (Tier 1)** | ✓ | ✓ | ✓ | ✓ | ✓ | ✓ | Insurable |
| **T-30 Bias (Tier 1)** | ✓ | ~ | ~ | ✓ | ~ | ✓ | Insurable; controls required |
| **T-08 Cost-DoS (Tier 2)** | ~ | ~ | ✓ | ~ | ~ | ~ | Silent / unpriced |
| **T-37 Shadow AI (Tier 2)** | ~ | ~ | ✓ | ~ | ~ | ✗ | Intentional employee act |
| **T-40 Lethal trifecta (Tier 4)** | ~ | ~ | ✓ | ✓ | ~ | ~ | Architectural controls |
| **T-43 AI-washing (Tier 4)** | ~ | ~ | ✓ | ✓ | ✗ | ✗ | Intentional act exclusion |
| **T-45 FM concentration (Tier 4)** | ✓ | ~ | ✗ | ✗ | ✗ | ✓ | Aggregation risk |

Note. The T-40 row scores current public-market pricing rather than theoretical insurability under a future configuration-rated form: trifecta posture can be underwritten as a risk selection condition even where current public products do not price the architectural peril itself. Criteria: (i) randomness/fortuity of loss; (ii) assessability of probability and severity; (iii) loss independence; (iv) maximum possible loss bounded; (v) economic feasibility; and (vi) absence of intentional act / moral hazard.

### *G. A note on adversarial attack characterization*

A subsidiary observation from the corpus, based on publicly available carrier materials only, is that some affirmative AI products appear to frame adversarial incidents against deployed AI systems as model performance or agent-failure events rather than as traditional cyber incidents. Public Munich Re materials [5, 6, 39] describe aiSure as a contractual performance guarantee structure addressing model underperformance; Armilla's public materials [16, 28] describe affirmative coverage for AI Model Error Liability and AI Agent Failures; AIUC's materials [17, 24] describe affirmative coverage for AI agent hallucination, IP infringement, and data leakage. By contrast, standard cyber wording continues to treat AI-mediated incidents under the unauthorized access framing, and Verisk CG 40 47's generative AI language may exclude adversarial perturbation or downstream injury to the extent the claim is framed as arising out of generative AI.

The structural implication is a hypothesis, not a claim about executed policy wording: if affirmative AI cover continues to frame prompt injection class events as performance stress events, those incidents may gradually move out of the silent-cyber debate and into AI-specific affirmative cover. Two counterarguments deserve weight against this hypothesis. First, the historical trajectory of cyber insurance is one of expansion rather than contraction: cyber lines have absorbed novel attack vectors (ransomware, business email compromise, dependency-chain compromise) rather than shedding them to adjacent markets, and there is no obvious reason prompt injection events should follow a different pattern. Second, the cause-of-loss in many adversarial AI incidents (compromised credentials, lateral movement, data exfiltration) maps cleanly onto cyber wording even when an LLM sits in the middle of the kill chain; the AI-as-instrumentality framing does not by itself displace cyber. The more likely near-term path is that prompt injection events remain primary cyber claims, with affirmative AI products acting as supplementary cover for the AI-specific elements (model error, hallucination, agent failure) that cyber wording does not naturally reach. We flag the migration hypothesis as a possibility worth tracking rather than as a prediction. We do not analyze executed policy wording here; any specific claim outcome depends on the contract, applicable law, and forum. The framing observation should be tested by future case law, policy form publication, and redacted claim experience.

Interpretation note. This research codes public product positioning and exclusion language, not final claim outcomes. Actual coverage depends on executed wording, jurisdiction, facts, retentions, sublimits, exclusions, and other insurance provisions. 'Affirmative' means public materials expressly identify the peril or use wording broad enough to capture it under this paper's decision rules; it does not mean every claim involving that peril would be paid. The full 55 by 26 matrix is reproduced in Appendix A.

## VII. SYSTEMIC AGGREGATION: FOUNDATION MODEL CONCENTRATION

Foundation model concentration (T-45) is the most analytically distinctive insurance economics question raised by the corpus. Enterprise AI deployment is concentrated among a small number of foundation model providers. Aon's Kalinich, in remarks reported directly by the Financial Times, articulated the underwriting problem: carriers may be able to absorb a large loss from a single insured, but cannot easily price systemic, correlated, aggregated risk from an upstream AI-provider failure [9]. We unpack the argument in three pieces.

### *A. Why this is not a standard correlated-loss problem*

Standard correlated-loss perils, such as a hurricane, a pandemic, or a power-grid outage, share two features that admit conventional reinsurance treatment: the trigger is exogenous (a named storm, a declared pandemic) and severity scales with measurable physical or epidemiological intensity. Foundation model concentration produces correlation through a different mechanism. A single upstream model behavior, such as a regression that introduces hallucination on a specific class of prompts, a security vulnerability disclosed in the model serving infrastructure, or a regulatory action that pulls a model from market, affects every cedent that deployed that model, simultaneously, with no exogenous trigger and no obvious severity scale. The closest analog is an upstream software supply-chain failure (e.g., the SolarWinds compromise of 2020), but software supply-chain failures have not historically produced insurance-market-defining aggregation events the way named storms have.

### *B. Why classical insurability criteria fail*

Applying the criteria from Section II.A to T-45: (i) the loss event itself is random, since foundation model failures are not predictable on a calendar and individual cedents do not select the failure date, so randomness is preserved; (ii) probability is partially assessable from upstream-vendor incident history (OpenAI, Anthropic, and Google all publish status pages and post-mortems), though base rates are sparse; (iii) loss independence fails sharply, because by construction every cedent using the failed upstream model experiences claim conditions simultaneously; (iv) maximum possible loss is unbounded in the limit because the cedent base is large and growing; (v) economic feasibility fails because a premium that priced the correlated maximum loss would exceed any plausible cedent willingness-to-pay; (vi) the moral hazard criterion is largely satisfied because cedents do not control upstream provider behavior. The binding constraints are (iii) loss independence and (v) economic feasibility, both of which fail for structural reasons.

### *C. Candidate market structures*

The relevant question is not merely which systemic risk templates exist, but which insurability constraint each template relaxes and why that structure has not yet formed for AI. Table VI evaluates six candidates against the criteria introduced in Section II.A and applied to T-45 in Section VII.B. The final column ("Seq.") gives the expected sequence of formation under current public market conditions, from earliest and most already visible (1) to latest and most contingent (6).

TABLE VI: Comparative evaluation of candidate structures for foundation model aggregation risk.

| Structure | Constraint relaxed | Trigger observable | Moral hazard | Feasibility / precedent | Seq. |
|---|---|---|---|---|---|
| **Provider indemnification / private ordering (not insurance)** | Partly bounds defense-cost uncertainty for IP/output claims; does not solve portfolio correlation. | Contractual claim under provider indemnity terms. | Low insurer moral hazard; depends on customer compliance with provider terms. | Visible in OpenAI, Anthropic, and Google Cloud terms [52-54]. Contractual indemnity, not insurance. | 1 |
| **Cedent diversification + reps/warranties** | Improves independence by reducing dependence on one upstream model provider. | SLA breach, regression notice, outage report, or failover test. | Favorable if second-source capacity is real rather than nominal. | Available now through procurement and architecture. | 2 |
| **Captive / RRG or mutual pool** | Improves feasibility by retaining attritional loss and buying excess tail protection. | Member claims within an AI-incident taxonomy. | Moderate; members control underwriting and loss-prevention standards. | RRGs are member-owned liability insurers [55]; plausible in homogeneous sectors. | 3 |
| **Reinsurance pool** | Addresses independence and feasibility at portfolio level. | Aggregate claims tied to a named upstream-provider incident. | Moderate; needs cedent controls and anti-selection rules. | Terrorism/pandemic-pool precedent; AI-using cedent pool is hard to define. | 4 |
| **Cat-bond / ILS** | Transfers tail aggregation risk to capital markets. | Parametric provider incident: outage, rollback, suspension, or claims threshold. | Low for cedents; high basis risk if trigger misses actual loss. | Strong natural-cat precedent; missing AI-specific trigger metric. | 5 |
| **Government backstop** | Solves tail feasibility when private capital cannot hold systemic maximum loss. | Statutory certification plus aggregate insured-loss threshold. | Highest; needs deductibles, retentions, and anti-free-riding rules. | TRIA-like precedent; political burden high. | 6 |

The ranking suggests a sequence rather than a single solution. Private ordering and cedent-side diversification are already forming because they do not require regulatory action and can be written into enterprise procurement. Captive or risk retention group structures are the next plausible private mechanism for sectors with homogeneous AI exposures, because they allow members to retain attritional AI loss while buying reinsurance above the pool. Reinsurance pools, ILS, and a public backstop become plausible only after the market has enough incident data to define a trigger and enough loss experience to persuade cedents, reinsurers, or governments that the correlation problem is material rather than hypothetical.

The same comparison also explains non-formation. Provider indemnities are narrow because upstream providers will not accept open-ended liability for every downstream deployment.

Cat bonds have not formed because there is no accepted parametric trigger: a provider outage, model rollback, or post-mortem may not correlate cleanly with insured loss. A TRIA-style backstop has the highest tail capacity but the weakest near-term political economy, because no AI aggregation event has yet forced a public-private solution. The most testable near-term prediction is therefore not a government program, but explicit aggregation sublimits, provider-event exclusions, or captive/RRG experimentation in sectors that depend heavily on a small number of foundation models.

### *D. Munich Re aiSure's aggregation-management posture*

Munich Re's aiSure is a contractual performance-guarantee arrangement under which the Insured offers a Performance Guarantee to its customers (Guarantee Holders) and Munich Re backs that Guarantee [5, 6, 39]. The product is volume-priced, and aggregate exposure is managed through per-Guarantee limits and reporting cadence. Without examining executed policy wording, we observe that the public structure has natural aggregation management features: volume-based pricing means premium scales with usage; per-Guarantee limits mean exposure is contained at the customer level; and ongoing reporting allows the carrier to identify a developing cluster of underperformance events early. These features do not solve the foundation model concentration problem at the market level, but they show why the earliest AI aggregation mechanisms are likely to appear through private ordering, contractual limits, reporting cadence, and reinsurance attachment points before they appear as a public backstop.

## VIII. IMPLICATIONS

### *A. For risk managers*

Risk managers procuring AI insurance should map their exposure against the four-tier frontier explicitly. The corpus suggests three procurement priorities. First, where a peril sits in Tier 1 (affirmatively covered), buyers should compare structures across products: the same peril (hallucination, T-04) is covered by Armilla as broad liability, by Munich Re aiSure as a performance guarantee structure, by AIUC as agent-failure liability, by Testudo as mid-market liability, and by Hiscox/Chaucer as warranty-style. The structural choice has cash-flow, duty-to-defend, and limit-stacking consequences. Second, where a peril sits in Tier 2 (silent-AI exposure), the buyer's existing cyber, tech E&O, D&O, EPLI, and crime policies should be re-papered explicitly: "arising out of AI" wording should be made clear in either direction. Verisk CG 40 47 is now adoptable by any US carrier using ISO forms, and one affirmative-side market participant has predicted broad adoption [27]; the matrix, however, codes Verisk narrowly where the public record supports direct threat capture. Third, Tier 4 perils should be addressed architecturally, contractually, and through governance discipline; insurance is at best a complement to those mechanisms, not a substitute.

### *B. For reinsurers*

The aggregation argument articulated by Aon's Kalinich [9] is the central reinsurance question. If foundation model concentration produces correlated losses across cedents, the standard-setting question for reinsurers is whether to (a) treat foundation model failure as a systemic peril analogous to a named-storm event in property treaty, with explicit aggregation limits and cat-bond-like coverage, or (b) exclude foundation model failure from underlying covers and require cedents to purchase standalone aggregation cover. Section VII evaluates six candidate structures; we suggest reinsurers and brokers conduct a Lloyd's-style market study on AI-aggregation triggers in the spirit of the 2019 to 2020 silent-cyber bulletins.

### *C. For regulators*

For US state insurance regulators, the rapid split between affirmative AI cover and active exclusion poses a market-conduct question: does the average mid-market buyer understand that, after the January 2026 rollout of ISO generative AI exclusions, its CGL policy may exclude AI losses unless it affirmatively buys cover? The NAIC Model Bulletin on AI has been adopted by more than 20 states plus DC by spring 2026 [23], but it addresses insurer governance of AI in underwriting and claims rather than coverage architecture for AI-mediated losses on the insured side. NY DFS Circular Letter No. 7 (July 2024) similarly addresses insurer-side AI use [23]. A useful next step would be a model bulletin or circular addressing affirmative-versus-silent-versus-excluded disclosure standards for AI cover, modeled on Lloyd's silent-cyber bulletins of 2019 to 2020.

For the EU, the AI Act's high risk system regime requires operators to assemble conformity assessment, technical documentation, transparency, and human oversight evidence on a phased timetable [46]. Affirmative carriers (Armilla, AIUC, Axa XL) already discuss defense-cost and regulatory-response cover in relation to AI regulation. Regulator-published alignment between conformity assessment evidence and insurance underwriting evidence would materially reduce duplication for cedents.

### *D. For the AI-security community*

The AI-security community publishes threat catalogs at high cadence: OWASP LLM Top 10, OWASP Agentic Top 10, OWASP Agentic Skills Top 10, MITRE ATLAS, AIVSS (in development). Insurance carriers reference these as underwriting evidence (AIUC-1, Armilla AI Product Warranty pricing, Munich Re HSB SMB underwriting). A standing liaison between the OWASP GenAI Security Project, MITRE, and the affirmative AI cover MGAs would help reduce the lag between catalog update and policy form update; at present, AST10 was published in March 2026 but no product in our corpus has yet refreshed wording explicitly for AST01 to AST10.

## IX. LIMITATIONS

Wording variability. Coverage in any specific claim depends on the executed policy wording, which varies by carrier, jurisdiction, broker negotiation, and renewal cycle. The matrix codes publicly marketed policy intent; final coverage in any given case is a question of the executed contract.

No analysis of contractual liability insurance policy (CLIP) wording. References to aiSure rely solely on Munich Re's publicly available materials [5, 6, 39]. The structural observations we draw from public marketing are hypotheses about policy design, not findings about executed contracts.

Surplus-lines and jurisdictional issues. Several products in the corpus are surplus-lines policies. Surplus-lines status materially affects the buyer's protection in carrier insolvency and may affect choice-of-law and forum questions. We do not analyze jurisdictional questions in depth.

Carrier filings. For carrier-specific exclusion filings (I-24), state-by-state forms are not always public. We rely on trade-press summaries triangulated across multiple sources where possible, with explicit Tier-C labeling.

Threat-catalog volatility. The OWASP Agentic Skills Top 10 (T-46 to T-55) remains an incubator project as of May 2026. The AST cluster is therefore the most volatile portion of the matrix.

No claim-frequency or severity data. We do not present claim-frequency or severity data because none of the affirmative AI products in our corpus has a published loss development triangle; Armilla and AIUC have written affirmatively for less than two policy years. Munich Re aiSure has a longer history but does not publish line-level loss data. Future work should pursue claim experience data, ideally via a regulator-mediated data call. A companion paper, "Coverage Without Casualties?" [58], takes a first step in this direction. It joins the 55-threat catalog of this paper to 3,054 publicly documented real-world AI incidents (2024 to 2026) and finds that affirmative coverage breadth bears little relation to observed incidence: model drift draws seven affirmative coverage cells with zero recorded incidents, while deepfake fraud, the most frequent recent peril, draws three. Incident data grounds threat prevalence and severity, not claim frequency or loss cost, so the limitation noted here remains open.

Single-interpretation coding. The matrix uses three independent human coders with LLM-assisted source review and group adjudication of disagreements (Section III.F), with inter-rater reliability statistics reported in supplementary material. Even with this design, each cell carries a single best evidence code rather than a distribution across plausible coverage arguments. A probabilistic coding that reflected coder confidence and source-tier weighting would be a useful extension. The per-cell evidence index and coder-vote record are published as supplementary material to enable replication and re-coding.

Temporal validity. This paper is dated May 6, 2026; the AI-insurance market changed dramatically in the prior 12 months and will likely change again. We are explicit that the matrix is a snapshot. The methodology and the four-tier frontier are intended as the durable contributions; the specific cell values are expected to evolve.

## X. CONCLUSION

We have constructed and analyzed a 55-threat by 26-product coverage matrix for AI-insurance market positioning as of May 2026. Four findings are central. First, the public AI-insurance market is consistent with bifurcation: affirmative AI products are expanding while exclusion endorsements narrow legacy forms. Second, affirmative coverage is specializing by peril rather than converging on a single general AI policy form. Third, prompt-injection-class events may migrate only partially from silent cyber into affirmative AI cover, with cyber likely remaining primary for many incidents and AI-specific policies supplementing model-error or agent-failure elements. Fourth, the Tier 4 boundary cases are structurally distinct: lethal-trifecta risk is first an architectural-control problem, AI-washing is a familiar intentional-act problem, and foundation model concentration is the genuinely novel portfolio-correlation problem.

Three contributions support these findings: a published codebook and decision-rule set that makes the matrix replicable; a four-tier insurability frontier that organizes the corpus against classical insurability criteria; and a focused treatment of foundation model concentration as the systemic aggregation case requiring a market structure beyond conventional liability cover. We invite future work that examines executed policy wording (subject to confidentiality), extends inter-rater reliability analysis and publishes per-coder vote distributions, and pursues regulator-mediated claim experience data.

The four-tier frontier yields specific predictions that the next 12 to 24 months can test. The bifurcation thesis would be confirmed by sustained growth in Tier 1 affirmative capacity alongside continued spread of CG 40 47/48-style exclusions, with the silent-AI exposure zone (Tier 2) shrinking as both ends advance; it would be weakened if affirmative capacity contracts under early loss experience, if exclusion adoption remains limited to narrow threat classes, or if a major loss event drives a return to broad legacy responsiveness in cyber and tech E&O. The peril-specialization thesis would be confirmed if the threat-to-market crosswalk in Table III hardens (Munich Re consolidating around drift, Apollo ibott around autonomous system liability, Coalition around deepfake response, etc.); it would be falsified if affirmative carriers converge to undifferentiated multi-peril forms. The Tier 4c thesis, that foundation model concentration is the binding insurability constraint, would be confirmed first by explicit aggregation sublimits, provider-event exclusions, or captive/RRG experimentation in sectors that depend heavily on a small number of foundation models; later-stage confirmation would include an AI-aggregation cat bond, a brokered reinsurance pool, or a TRIA-style backstop discussion. It would be falsified if a meaningful upstream-provider failure produced widespread cedent loss without significant insurance market disruption, suggesting that the correlation problem is smaller than its underwriting profile suggests. The durable contributions of this paper are intended to be the methodology and the four-tier framework; specific cell values and named market positions are expected to evolve, and we publish the matrix in a form designed to support recoding as the market changes.

## DATA AND REPRODUCIBILITY

The full 55-threat by 26-product matrix, source-citation index, per-cell evidence pointers, and threat-to-market crosswalk are reproduced in Appendix A. All URL-based sources were verified accessible as of May 6, 2026.

## DISCLOSURE

This paper presents an analytical mapping of publicly marketed insurance policies and exclusion endorsements against published threat taxonomies. It does not constitute legal advice, coverage advice, or investment advice. Risk managers and counsel should obtain executed policy wording and qualified advice for any specific placement decision. The authors have no financial interest in any of the carriers, MGAs, or underwriting facilities discussed. The paper does not analyze, quote, or rely on confidential or executed policy wording from any carrier; references to specific products rely solely on publicly available carrier materials and public record filings.

## REFERENCES

[1] OWASP GenAI Security Project. OWASP Top 10 for LLM Applications 2025. https://genai.owasp.org/llm-top-10/

[2] OpenAI. Continuously hardening ChatGPT Atlas against prompt injection. 22 December 2025. https://openai.com/index/hardening-atlas-against-prompt-injection/

[3] OWASP GenAI Security Project. OWASP Top 10 for Agentic Applications 2026 (ASI01 to ASI10). 9 December 2025.

[4] MITRE Corporation. MITRE ATLAS v5.1.0. November 2025. https://atlas.mitre.org/

[5] Munich Re. Munich Re Generating content with AI: An IP-infringement minefield?. 2024 to 2025. https://www.munichre.com/en/solutions/for-industry-clients/insure-ai/ai-whitepaper.html

[6] Munich Re. aiSure: More AI Opportunity. Less AI Risk. https://www.munichre.com/en/solutions/for-industry-clients/insure-ai.html

[7] Willison, S. The lethal trifecta for AI agents. 16 June 2025. https://simonwillison.net/2025/Jun/16/the-lethal-trifecta/

[8] Berliner, B. Limits of Insurability of Risks. Prentice-Hall, 1982.

[9] Harris, L., & Criddle, C. Insurers retreat from AI cover as risk of multibillion-dollar claims mounts. Financial Times, 22 November 2025. https://www.ft.com/content/abfe9741-f438-4ed6-a673-075ec177dc62

[10] World Economic Forum. Global Cybersecurity Outlook 2026. 12 January 2026. https://reports.weforum.org/docs/WEF_Global_Cybersecurity_Outlook_2026.pdf

[11] OWASP Foundation. OWASP Agentic Skills Top 10 (incubator project). https://owasp.org/www-project-agentic-skills-top-10/

[12] Beurer-Kellner et al. (IBM, Invariant Labs, ETH Zurich, Google, Microsoft). Design Patterns for Securing LLM Agents against Prompt Injections. 2025. https://arxiv.org/abs/2506.08837

[13] Liaghati, C. (MITRE / NIST CSRC). MITRE ATLAS Overview. September 2025.

[14] Google Cloud Threat Intelligence Group (Mandiant). Widespread Data Theft Targets Salesforce Instances via Salesloft Drift (UNC6395). August 2025. https://cloud.google.com/blog/topics/threat-intelligence/data-theft-salesforce-instances-via-salesloft-drift

[15] FINRA. Cybersecurity Alert: Salesloft Drift AI Supply Chain Attack. September 2025.

[16] Armilla AI. Armilla AI Insurance, Lloyd's Coverholder. https://www.armilla.ai/ai-insurance

[17] AIUC. AIUC-1: AI Agent Standard. https://www.aiuc-1.com/

[18] Knostic AI. Privacy guardrails for GenAI (Copilot/Glean). https://www.knostic.ai/

[19] Pearcy, S. (HiddenLayer Research). The Lethal Trifecta and How to Defend Against It. 25 November 2025. https://www.hiddenlayer.com/research/the-lethal-trifecta-and-how-to-defend-against-it

[20] Hunton Insurance Recovery Blog. Affirmative Artificial Intelligence Insurance Coverages Emerge. May 2025.

[21] Independent Agent / Big I. Verisk to Roll Out New General Liability Exclusions for Generative AI Exposures. October 2025. https://www.independentagent.com/vu_resource/verisk-to-roll-out-new-general-liability-exclusions-for-generative-ai-exposures/

[22] Gridex. Verisk CG 40 47: What the New AI Exclusions Mean for Your Commercial Clients. March 2026.

[23] National Association of Insurance Commissioners. Insurance Topics: Artificial Intelligence; AI Model Bulletin State Adoption Map. April 2026. https://content.naic.org/insurance-topics/artificial-intelligence.

[24] Sharon Goldman (Fortune). Exclusive: Who covers the damage when an AI agent goes rogue? This startup has an insurance policy for that. 23 July 2025. https://fortune.com/2025/07/23/ai-agent-insurance-startup-aiuc-stealth-15-million-seed-nat-friedman/

[25] U.S. Securities and Exchange Commission. SEC Charges Two Investment Advisers with Making False and Misleading Statements About Their Use of Artificial Intelligence. March 2024. https://www.sec.gov/newsroom/press-releases/2024-36

[26] FFNews / FFInsurtech. Armilla AI Raises Lloyd's-Backed Coverage to $25M as Traditional Insurers Retreat. January 2026.

[27] Reinsurance News. Testudo launches AI insurance underwriting platform backed by Lloyd's Lab. June 2025. https://www.reinsurancene.ws/testudo-launches-ai-insurance-underwriting-platform-backed-by-lloyds-lab/

[28] Armilla. Armilla Launches Affirmative AI Liability Insurance with Chaucer. 30 April 2025.

[29] Munich Re HSB. HSB Introduces AI Liability Insurance for Small Businesses. 18 March 2026. https://www.munichre.com/hsb/en/products/services/artificial-intelligence-insurance.html

[30] Coalition. Coalition Adds Deepfake Response Endorsement to its Cyber Insurance Policies Globally. 9 December 2025. https://www.coalitioninc.com/announcements/coalition-adds-deepfake-response-endorsement

[31] Apollo Group / Marsh. Marsh and Apollo's ibott develop first-of-its-kind insurance facility for Uber to accelerate autonomous ride-hailing. 12 March 2026.

[32] Vouch. Insurance for AI Companies (Vouch + Hiscox / Corix). https://www.vouch.us/technology/ai

[33] Business Insurance. Insurers, brokers adjust as AI exclusions emerge (Coalition, Axa XL endorsements). April 2026.

[34] FinancialContent / Risk Specialty Group. Insurance Carriers Add AI Exclusions to Design Professional E&O Policies. January 2026.

[35] American Banker. Insurers likely to exclude gen AI, startups wait in wings (Testudo, Vouch, Armilla). November 2025.

[36] S&P Global Market Intelligence. As insurers retreat from AI risk, one startup plans to fill the gap. February 2026.

[37] Insurance Intel Substack. Chubb Is Excluding the Risk Its Own CEO Says AI Will Solve (W. R. Berkley PC 51380). March 2026.

[38] The Insurer / Slipcase. Insurer in Full: US liability insurers explore AI exclusions. October 2025.

[39] Mosaic Insurance. Transactional liability, partnership with Munich Re aiSure. https://www.mosaicinsurance.com/resources/press-releases/~/mosaic-partners-with-munich-res-aisure-to-provide-pioneering-coverage-for-ai-vendors/

[40] IBM X-Force. What OpenClaw reveals about agentic AI security risks (ClawHavoc, 1,184 malicious skills). April 2026.

[41] Antiy CERT. ClawHavoc: Analysis of Large-Scale Poisoning Campaign Targeting the OpenClaw Skill Market for AI Agents. 6 February 2026. https://www.antiy.net/p/clawhavoc-analysis-of-large-scale-poisoning-campaign-targeting-the-openclaw-skill-market-for-ai-agents/

[42] Aim Security / Microsoft. EchoLeak (CVE-2025-32711), first zero-click prompt injection in Microsoft 365 Copilot. 2025.

[43] OX Security. The Mother of All AI Supply Chains: Critical, Systemic Vulnerability at the Core of Anthropic's MCP. April 2026. https://www.ox.security/blog/the-mother-of-all-ai-supply-chains-critical-systemic-vulnerability-at-the-core-of-the-mcp/

[44] Meta AI Research. LlamaFirewall: An open source guardrail system for building secure AI agents. April 2025.

[45] Wang & Li. OpenGuardrails: A Configurable, Unified, and Scalable Guardrails Platform for Large Language Models. October 2025. https://arxiv.org/abs/2510.19169v2

[46] European Commission. Timeline for the Implementation of the EU AI Act. AI Act Service Desk. https://ai-act-service-desk.ec.europa.eu/en/ai-act/timeline/timeline-implementation-eu-ai-act

[47] Reinsurance News. Insurers expected to introduce GenAI liability exclusions: Evercore ISI. December 2025.

[48] Traverse Legal. AI Insurance Requirements: Insurance May Not Cover Your AI Failures (D&O / E&O / EPLI exclusions detail). April 2026.

[49] Romanosky, S., Ablon, L., Kuehn, A., & Jones, T. Content Analysis of Cyber Insurance Policies: How Do Carriers Price Cyber Risk? Journal of Cybersecurity, 5(1), 2019.

[50] Talesh, S. A. Data Breach, Privacy, and Cyber Insurance: How Insurance Companies Act as "Compliance Managers" for Businesses. Law & Social Inquiry, 43(2), 417 to 440, 2018.

[51] Baker, T. & Shortland, A. The Government Behind Insurance Governance: Lessons for Ransomware. Regulation & Governance, October 2023. https://doi.org/10.1111/rego.12505. See also Baker, T. & Logue, K. D., & Saiman, C. Insurance Law and Policy: Cases and Materials, 5th ed., Wolters Kluwer, 2021, ch. 1.

[52] OpenAI. Service Terms. 9 January 2026. https://openai.com/policies/service-terms/

[53] Anthropic. Expanded legal protections and improvements to our API. 19 December 2023. https://www.anthropic.com/news/expanded-legal-protections-api-improvements; see also Anthropic, Updates to Consumer Terms and Privacy Policy. 28 August 2025. https://www.anthropic.com/news/updates-to-our-consumer-terms (noting that the 2025 consumer updates do not apply to commercial terms, Claude for Work, Claude for Government, Claude for Education, or API use).

[54] Google Cloud. Service Specific Terms, current version, Generative AI Services indemnification; and Google Cloud Generative AI Indemnified Services. https://cloud.google.com/terms/service-terms and https://cloud.google.com/terms/generative-ai-indemnified-services

[55] National Association of Insurance Commissioners. Insurance Topics: Risk Retention Groups. https://content.naic.org/insurance-topics/risk-retention-groups

[56] Hansen R. and Shokrai M. (Google Cloud). Expanding our Risk Protection Program with new insurance partners and AI coverage. 16 May 2025. https://cloud.google.com/blog/products/identity-security/whats-new-with-google-clouds-risk-protection-program/. See also Risk Protection Program product page: https://cloud.google.com/security/products/risk-protection-program

[57] Y. T. Shen, K. Toyoda, and A. Leung, "MCP-38: A Comprehensive Threat Taxonomy for Model Context Protocol Systems," arXiv preprint arXiv:2603.18063, 2026. [Online]. Available: https://arxiv.org/abs/2603.18063

[58] Leung, A., Zhang, R., Toyoda, K., & Loh, S. Coverage Without Casualties? An Empirical Test of the AI Insurability Frontier Against Recent Incident Evidence. AIFT, 8 June 2026. https://vulcanlab.github.io/genai-security-incidents/

## APPENDIX A: FULL 55 × 26 MATRIX

TABLE A1: Threat × insurance-product coverage status. A = affirmative; S = silent / gray; X = excluded; · = no public coverage position.

| Threat | Description | I-01 | I-02 | I-03 | I-04 | I-05 | I-06 | I-07 | I-08 | I-09 | I-10 | I-11 | I-12 | I-13 | I-14 | I-15 | I-16 | I-17 | I-18 | I-19 | I-20 | I-21 | I-22 | I-23 | I-24 | I-25 | I-26 |
|---|---|---|---|---|---|---|---|---|---|---|---|---|---|---|---|---|---|---|---|---|---|---|---|---|---|---|---|
| T-01 | **Prompt Injection** | A | A | A | A | · | · | S | S | · | · | · | · | · | · | A | S | · | · | · | · | X | · | · | X | X | S |
| T-02 | **Indirect Prompt Injection** | A | A | A | A | · | S | · | · | · | · | · | A | · | · | · | S | S | · | · | · | · | X | · | · | X | · |
| T-03 | **Sensitive Information Disclosure** | A | A | A | A | · | A | · | A | · | A | · | · | · | · | A | S | · | · | · | · | · | X | · | · | X | · |
| T-04 | **Hallucination** | A | A | A | A | S | A | · | · | · | A | A | A | · | · | · | S | S | S | · | · | · | X | · | · | X | S |
| T-05 | **Insecure Output Handling** | A | · | A | · | · | A | · | · | · | · | · | · | · | · | A | S | · | · | · | · | · | X | · | · | X | · |
| T-06 | **System Prompt / Hidden Context Exposure** | A | A | A | A | · | A | · | · | · | A | · | · | · | · | S | S | · | · | · | · | · | X | · | · | X | · |
| T-07 | **Vector / RAG Poisoning** | A | A | A | A | · | A | · | · | · | A | · | · | · | · | S | S | · | · | · | · | · | X | · | · | X | · |
| T-08 | **Cost Denial-of-Service** | · | · | · | · | S | S | · | · | · | · | · | · | · | · | S | S | · | · | · | · | · | · | · | · | · | · |
| T-09 | **Jailbreak / Safety Bypass** | A | A | A | A | · | A | · | · | · | A | · | · | · | · | S | S | · | · | · | · | · | X | · | · | X | · |
| T-10 | **IP Infringement** | A | A | A | A | · | · | · | A | A | A | · | · | A | · | · | A | S | · | · | · | · | X | X | · | · | S |
| T-11 | **Goal Hijacking** | A | · | A | A | · | A | · | · | · | A | · | · | · | A | A | S | · | · | · | · | · | X | · | · | X | · |
| T-12 | **Tool Misuse / Excessive Agency** | A | · | A | A | · | A | · | · | · | A | · | · | · | A | A | S | · | · | · | · | · | X | · | · | X | · |
| T-13 | **Identity and Privilege Abuse** | A | · | A | · | · | A | · | · | · | · | · | · | · | A | A | S | · | · | · | · | · | X | · | · | X | · |
| T-14 | **Supply Chain Compromise** | A | · | A | · | · | A | · | · | · | · | · | · | · | · | A | S | · | · | · | · | · | X | · | · | X | · |
| T-15 | **Unexpected Code Execution** | A | · | A | · | · | A | · | · | · | · | · | · | · | A | A | S | · | · | · | · | · | X | · | · | X | · |
| T-16 | **Memory / State Poisoning** | A | · | A | A | · | A | · | · | · | · | · | · | · | A | A | S | · | · | · | · | · | X | · | · | X | · |
| T-17 | **Insecure Inter-Agent Communication** | A | · | A | · | · | A | · | · | · | · | · | · | · | A | A | S | · | · | · | · | · | X | · | · | X | · |
| T-18 | **Cascading Failure** | A | · | A | A | · | A | · | · | · | A | · | · | · | A | A | S | · | · | · | · | · | X | · | · | X | · |
| T-19 | **Trust Exploitation** | A | · | A | A | · | A | · | · | · | A | · | · | · | A | A | S | · | · | · | S | · | X | · | · | X | · |
| T-20 | **Rogue Agents** | A | · | A | A | · | A | · | · | · | · | · | · | · | A | · | S | S | · | · | · | · | X | X | · | X | · |
| T-21 | **MCP Server Compromise** | · | · | S | S | · | A | · | · | · | · | · | S | · | S | A | S | · | · | · | · | · | X | · | · | · | · |
| T-22 | **Unauthorized MCP Server Exposure** | · | · | A | S | · | A | · | · | · | · | · | S | · | · | A | S | · | · | · | · | · | X | · | · | X | · |
| T-23 | **MCP Context Manipulation** | · | · | A | S | · | S | · | · | · | · | · | · | · | · | A | S | · | · | · | · | · | X | · | · | · | · |
| T-24 | **OAuth / Token Cascade** | A | · | A | S | · | A | · | · | · | · | · | S | · | A | A | S | · | · | · | · | · | X | · | · | X | · |
| T-25 | **Model Evasion / KYC Bypass** | A | A | A | A | · | A | S | A | · | A | · | · | · | A | A | S | · | · | · | S | · | X | · | · | X | · |
| T-26 | **Data Poisoning** | A | A | A | A | · | A | · | A | · | · | A | · | · | · | · | A | S | · | · | · | · | · | X | · | · | S |
| T-27 | **Model Extraction** | A | · | A | · | · | A | · | · | · | A | · | · | · | · | A | A | · | · | · | · | · | X | · | · | X | · |
| T-28 | **Model Inversion / Membership Inference** | A | · | A | A | · | A | · | · | · | A | · | · | · | · | A | S | · | · | · | · | · | X | · | · | X | · |
| T-29 | **Model Drift** | A | A | A | A | S | · | · | · | · | A | A | A | · | · | · | S | · | · | · | · | · | X | X | · | X | · |
| T-30 | **Bias and Fairness Failures** | A | A | A | S | · | · | · | · | · | A | · | A | · | · | · | S | S | S | · | · | · | X | X | · | X | · |
| T-31 | **Compromised Foundation Model** | A | · | A | · | S | A | · | · | · | · | · | · | · | · | A | S | · | · | · | · | · | X | · | · | X | · |
| T-32 | **Insecure AI Infrastructure** | · | · | A | · | · | A | · | · | · | · | · | · | · | · | A | S | · | · | · | · | · | X | · | · | X | · |
| T-33 | **Lack of AI Bill of Materials (No AI-BOM)** | · | · | S | · | · | S | · | · | · | · | · | · | · | · | · | · | · | · | · | · | · | · | · | · | · | · |
| T-34 | **Insecure AI-Generated Code** | A | · | A | S | · | A | · | · | · | A | · | · | · | · | · | A | S | · | · | · | · | · | X | · | · | S |
| T-35 | **Deepfake Fraud** | · | · | · | · | · | A | A | · | · | S | · | · | · | · | · | A | S | · | · | · | · | X | · | · | · | · |
| T-36 | **AI-Enhanced Phishing** | · | · | · | · | · | A | · | · | · | · | · | · | · | S | A | S | · | · | · | · | · | X | · | · | · | · |
| T-37 | **Shadow AI** | · | · | · | · | · | · | · | · | · | · | · | · | · | S | S | · | · | · | · | · | X | · | · | X | · | · |
| T-38 | **AI Offensive Operations** | · | · | · | · | · | A | · | · | · | · | · | · | · | A | A | · | · | · | · | · | · | · | · | · | · | · |
| T-39 | **Regulatory Non-Compliance** | A | · | A | A | · | · | · | A | · | A | · | · | A | · | · | · | · | · | A | · | X | X | · | X | · | · |
| T-40 | **Lethal Trifecta** | S | S | S | S | · | S | · | · | · | S | · | · | · | · | · | S | S | S | S | · | S | · | · | · | · | · |
| T-41 | **Browser-Agent Hijacking** | · | · | A | A | · | A | · | · | · | · | · | · | · | A | A | S | · | · | · | · | · | X | · | · | X | · |

| Threat | Description | I-01 | I-02 | I-03 | I-04 | I-05 | I-06 | I-07 | I-08 | I-09 | I-10 | I-11 | I-12 | I-13 | I-14 | I-15 | I-16 | I-17 | I-18 | I-19 | I-20 | I-21 | I-22 | I-23 | I-24 | I-25 | I-26 |
|---|---|---|---|---|---|---|---|---|---|---|---|---|---|---|---|---|---|---|---|---|---|---|---|---|---|---|---|
| T-42 | **Tool Description Injection** | · | · | A | S | · | A | · | · | · | · | · | S | · | · | A | S | · | · | · | · | · | X | · | · | · | · |
| T-43 | **AI-Washing** | · | · | · | · | · | · | · | · | · | · | · | · | · | · | · | S | X | · | X | · | · | · | · | · | · | · |
| T-44 | **AI Worms** | A | · | A | · | · | A | · | · | · | · | · | · | · | A | A | S | · | · | · | · | · | X | · | · | X | · |
| T-45 | **Foundation Model Concentration** | · | · | S | · | S | · | · | · | · | · | · | · | · | S | S | · | · | · | · | · | · | · | · | · | · | · |
| T-46 | **Malicious AI Skills / Plugins** | · | · | A | A | · | A | · | · | · | · | · | S | · | · | A | A | S | · | · | · | · | · | X | · | · | S |
| T-47 | **Skill Supply Chain Compromise** | · | · | A | A | · | A | · | · | · | · | · | S | · | · | A | A | S | · | · | · | · | · | X | · | · | S |
| T-48 | **Over-Privileged AI Skills** | · | · | A | S | · | A | · | · | · | · | · | S | · | · | A | A | S | · | · | · | · | · | X | · | · | S |
| T-49 | **Insecure Skill Metadata** | · | · | A | S | · | A | · | · | · | · | · | S | · | · | · | A | S | · | · | · | · | · | X | · | · | · |
| T-50 | **Unsafe Skill Deserialization** | · | · | A | S | · | A | · | · | · | · | · | · | · | · | A | A | S | · | · | · | · | · | X | · | · | S |
| T-51 | **Weak Skill Isolation** | · | · | A | S | · | A | · | · | · | · | · | · | · | A | A | S | · | · | · | · | · | X | · | · | X | · |
| T-52 | **Skill Update Drift** | · | · | A | S | · | A | · | · | · | · | · | · | · | A | A | S | · | · | · | · | · | X | · | · | X | · |
| T-53 | **Poor Skill Scanning** | · | · | S | S | · | S | · | · | · | · | · | · | · | · | · | · | · | · | · | · | · | · | · | · | · | · |
| T-54 | **Lack of Skill Governance** | · | · | A | S | · | A | · | · | · | · | · | · | · | · | A | S | · | S | · | · | · | X | · | · | X | · |
| T-55 | **Cross Platform Skill Exploitation** | · | · | A | S | · | A | · | · | · | · | · | · | · | · | A | S | · | · | · | · | · | X | · | · | X | · |

Source: authors' analysis based on carrier product disclosures, Verisk filings, and trade-press reporting. See supplementary material for per-cell evidence pointers.

Column abbreviations. I-01 = Armilla AI Liability; I-02 = Munich Re aiSure; I-03 = AIUC AI Liability; I-04 = Testudo AI Liability; I-05 = Google Cloud Risk Protection Program (Beazley, Chubb, Munich Re; AI insurance endorsement May 2025) [56]; I-06 = Coalition AI cyber endorsement; I-07 = Coalition deepfake response; I-08 = Axa XL generative-AI endorsement; I-09 = HSB AI Liability for SMB; I-10 = Vouch + Hiscox/Corix; I-11 = Armilla AI Product Warranty; I-12 = Hiscox/Chaucer hallucination liability; I-13 = Tokio Marine Kiln / CFC IP defense; I-14 = Apollo ibott autonomous / agentic liability; I-15 = standard cyber; I-16 = technology E&O; I-17 = D&O; I-18 = EPLI; I-19 = crime/social engineering; I-20 = media liability; I-21 = Verisk generative-AI exclusions; I-22 = W. R. Berkley PC 51380; I-23 = Berkley CGL/multi-line exclusion; I-24 = carrier-specific multi-line filings; I-25 = design-professional E&O exclusions; I-26 = expected industry GenAI liability exclusions (Tier-C signals coded S, not X, absent filed wording).

Coding notes. I-05 is coded with five S cells (T-04, T-08, T-29, T-31, T-45) on Tier-A material that names AI-workload-level coverage but does not name threats individually; the Google Cloud 2023 IP indemnification [54] is contractual and not coded as insurance under I-05. I-26 is a forecast / Tier-C expected-exclusion column, so affected cells are coded S rather than X unless filed wording becomes public. T-33 remains S under I-03 and I-06 despite provenance and accountability references because public materials gesture at accountability or SBOM-like controls but do not affirmatively name lack of provenance / no-AI-BOM as a covered peril. T-45 S cells are per-policy silent positions only and do not represent aggregation-level coverage across cedents.